\documentclass[12pt]{article}
\usepackage{amsfonts,amssymb}

\def\C{\mathbb{C}}
\def\Q{\mathbb{Q}}
\def\R{\mathbb{R}}
\def\Z{\mathbb{Z}}

\def\bq{ \begin{equation} }
\def\eq{ \end{equation} }
\def\ben{ \begin{eqnarray} }
\def\en{ \end{eqnarray} }

\def\frac#1#2{{#1\over #2}}

\def\on#1#2{\mathop{\vbox{\ialign{##\crcr\noalign{\kern2pt}
$\scriptstyle{#2}$\crcr\noalign{\kern2pt\nointerlineskip}
\kern-2pt$\hfil\displaystyle{#1}\hfil$\crcr}}}\limits}

\newcommand{\spa}[3]{({\bf #1}, {\bf #2}\times {\bf #3})}

\title{{\bf Integrable quadratic Hamiltonians on \\ $so(4)$ and $so(3,1)$}}
\author{Vladimir V Sokolov\\
Landau Institute for Theoretical Physics, \\
Moscow, Russia, {\tt sokolov@itp.ac.ru}\vspace{1cm}\\
Thomas Wolf\\ Department of Mathematics,
Brock University\\ 
Ontario, Canada, {\tt twolf@brocku.ca}}

\begin{document}

\maketitle
\begin{abstract}
We investigate a special class of quadratic Hamiltonians on $so(4)$ and 
$so(3,1)$ and describe Hamiltonians that have additional polynomial 
integrals. One of the main results is a new integrable case 
with an integral of sixth degree. \vspace{1cm}\vspace{5mm} 
\end{abstract}

\section{Introduction}
In this paper we consider the following family of Poisson brackets
\begin{equation} \label{puas}
\{M_{i},M_{j}\}=\varepsilon_{ijk}\,M_{k}, \qquad
\{M_{i},\gamma_{j}\}=\varepsilon_{ijk}\,\gamma_{k}, \qquad
\{\gamma_{i},\gamma_{j}\}=\kappa \, \varepsilon_{ijk}\,M_{k} .
\end{equation}
Here $M_{i}$ and $\gamma_{i}$ are components of 3-dimensional vectors
$\bf{M}$ and $\bf{\Gamma}$, $\varepsilon_{ijk}$ is the totally skew-symmetric
tensor, $\kappa$ is a parameter. It is well-known that any linear Poisson
bracket is defined by an appropriate Lie algebra. The cases $\kappa=0,$
$\kappa>0$ and $\kappa<0$ correspond to the Lie algebras
$e(3)$, $so(4)$ and $so(3,1)$. In this paper we assume $\kappa\ne 0.$ 

Bracket (\ref{puas}) has the two Casimir functions
$$
J_{1}=({\bf M}, {\bf \Gamma}), \qquad  J_{2}=\kappa \vert \bf{M}\vert ^{2}
+\vert {\bf \Gamma}\vert ^{2},
$$
where $(\cdot,\cdot)$ stands for the standard dot product in $\R^{3}$. 
Hence,  for the Liouville integrability of the equations of
motion only one additional integral functionally independent of the
Hamiltonian and the Casimir functions is necessary.

The simplest nontrivial class of quadratic homogeneous
Hamiltonians of the form
\begin{equation}
H=({\bf M}, \, A {\bf M})+({\bf M}, \, B {\bf \Gamma})+({\bf \Gamma}, \, C {\bf \Gamma}),
\label{HAMHOM}
\end{equation}
where $A,B$ and $C$ are constant $3\times 3$-matrices,
has many important applications in rigid body dynamics. We call
Hamiltonians (\ref{HAMHOM}) having an additional polynomial integral
{\it integrable}. We say that (\ref{HAMHOM}) possesses an additional
integral of degree $k$ if
there is no non-trivial integral of degree less than $k$.
In this paper all coefficients of both  the Hamiltonian and the additional 
integral are supposed to be real constants. 

The case where the Hamiltonian (\ref{HAMHOM}) has a linear additional
integral of motion has been investigated by Poincare \cite{puan}.

There are two classical integrable cases, one 
found by Frahm-Schottky and one by Steklov,  
where the additional integral of motion is of second degree. It was proved in
\cite{winter} that any Hamiltonian (\ref{HAMHOM}) possessing an additional
second degree integral is equivalent to one of these two cases.

In 1986 Adler-van Moerbeke \cite{admer} and
Reyman-Semenov-tian Shansky \cite{reysem}
independently found a Hamiltonian of the form (\ref{HAMHOM}) with fourth
degree additional integral.

For the Frahm-Schottky, Steklov, and Adler-van
Moerbeke-Reyman-Semenov-tian Shansky cases all matrices $A,B$ and $C$
are diagonal. This special subclass of "diagonal" Hamiltonians
(\ref{HAMHOM}) was investigated by many authors but no new integrable
cases were found.  Probably only these three integrable cases exist
among "diagonal" Hamiltonians.

In the paper \cite{sok} the first integrable
"non-diagonal" Hamiltonian (\ref{HAMHOM}) with $\kappa=0$
(i.e. a  Hamiltonian of the Kirchhoff type describing the motion of a rigid
body in ideal fluid) was found. This Hamiltonian has a fourth degree additional
integral. A generalization of this Hamiltonian to the case
$\kappa\ne 0$  was reported
in \cite{bomasok}.
In the paper \cite{soktsig} the Hamiltonian has been rewritten in the form
\begin{equation}\label{genham}
H=({\bf M},\, A {\bf M})+\spa{b}{M}{\Gamma},
\end{equation}
where $A$ is a constant symmetric matrix, ${\bf b}\ne 0$ is a constant vector
and $\times$ stands for the skew product. It turns out that this class 
is very rich in integrable cases. In the paper \cite{sok2} all
Hamiltonians (\ref{genham}) with a quartic additional integral were described.
Moreover, it was mentioned in \cite{tsig,sok2,tsiggor} that the general Sklyanin
brackets \cite{sklyan} for the $XXX$-magnetic model lead to some
integrable Hamiltonians of the same kind.

The goal of our paper is a systematic investigation of 
integrable real Hamiltonians
(\ref{genham}) and their inhomogeneous generalizations
\begin{equation}\label{genhamnon}
H=({\bf M},\, A {\bf M})+({\bf b}, \, {\bf M}\times {\bf \Gamma})+
({\bf k},\, {\bf M})+({\bf n},\, {\bf \Gamma}),
\end{equation}
where ${\bf k}$ and ${\bf n}$ are constant vectors.

It is remarkable that all known Hamiltonians
(\ref{genham}) possessing additional polynomial integrals have also some
{\bf linear} partial integrals
\begin{equation}\label{palin}
P=({\bf u},\,{\bf M})+({\bf v},\,{\bf
\Gamma}),
\end{equation}
where ${\bf u}$ and ${\bf v}$ are constant vectors. Namely, for
some constant vectors ${\bf p}$ and ${\bf q}$ the following relation
\begin{equation}\label{parint}
\{H,\, P\}=\Big[({\bf p},\,{\bf M})+({\bf q},\,{\bf \Gamma})\Big]\cdot P 
\end{equation}
holds. 
This implies that the corresponding equations of motion
preserve the constraint $P=0$. These linear partial integrals turn out
to be factors of the polynomial integrals for all homogeneous 
Hamiltonians considered in this paper.

Because of this reason we start our study with a subsection 2.1 devoted to 
linear partial integrals. An interesting subclass of one-parametric families 
of Hamiltonians (\ref{genham}) arises there. In the next subsection 2.2 
we apply the Kowalewski-Lyapunov test, which is well-known in the 
Painleve analysis, to find all possibly integrable families 
from this subclass. Because of a continuous parameter in the Hamiltonian 
this test becomes extremely efficient. 
The main result of this consideration is a new 
integrable Hamiltonian on $so(3,1)$ with an additional sixth degree integral. 

In section 3 we are dealing with inhomogeneous Hamiltonians (\ref{genhamnon}).
As we claim in section 4, there are no Hamiltonians of the form 
(\ref{genham}) having additional integrals of degrees from 1 to 8 
other than examples described in section 2.  We have 
also verified that we found in section 3 {\it all} Hamiltonians 
(\ref{genhamnon}) having additional integrals of degrees from 1 to 6.

All computations for the paper have been done by the specialized
computer algebra package {\sc Crack}.  It is designed to solve
overdetermined polynomial differential and algebraic systems with an
emphasis on extremely large problems.  A major concern for any
operations performed by the program is the complexity of resulting
expressions.  In \cite{wolf} an overview of the package and examples
for its use in the classification of integrable systems are given.

In this paper we present integrable Hamiltonians and the corresponding 
additional integrals in a general vector form, which is invariant with 
respect to the orthogonal transformations. For computations with 
vectorial expressions extra code was written.

\section{Homogeneous integrable cases}

\subsection{Linear partial integrals}

In this section we describe all Hamiltonians of
the form (\ref{genham}) having linear partial integrals 
(\ref{palin}). Relation (\ref{parint}) is equivalent to a 
system of bi-linear algebraic equations 
for coefficients of the Hamiltonian and components of vectors 
${\bf u}$, ${\bf v}$, ${\bf p}$ and ${\bf q}$. 
Below we present the result of our investigation of this system.

One can check that there are two different possibilities, either case 1:
${\bf v}={\bf q}=0$, or case 2: vector ${\bf q}$  is equal to vector 
${\bf b}$ from formula (\ref{genham}).
\paragraph*{Case 1:}
Calculations show that any Hamiltonian having a linear partial 
integral with ${\bf v}={\bf q}=0$ belongs to a class
of "vectorial" Hamiltonians of the form
\begin{equation}\label{vecGEN}
H=c_{1} \Big({\bf a},\,{\bf b}\Big)\vert {\bf M}\vert ^{2}+c_{2}
\Big({\bf a},\,{\bf M}\Big)\Big({\bf b},\,{\bf M}\Big)+
\Big({\bf b},\, {\bf M}\times {\bf \Gamma}\Big),
\end{equation}
where ${\bf b}$ and ${\bf a}$ are constant vectors,
$c_{i}$ are constant scalars. In this case ${\bf u}={\bf b}$ and
${\bf p}=c_{2}\,{\bf a}\times {\bf b}.$ 
In other words, for Hamiltonian (\ref{vecGEN}) we have
\begin{equation}\label{pin}
\{H,\, ({\bf b},\,{\bf M})\}=
c_{2}({\bf a}\times {\bf b},\, {\bf M} )\cdot ({\bf b},\,{\bf M}).
\end{equation}
In the next subsection we will present four different integrable Hamiltonians of the form
(\ref{vecGEN}) that possess additional integrals of degrees 1, 3, 4 and 6.
\paragraph*{Case 2:}
In the case ${\bf q}={\bf b}$ the following conditions
$$
({\bf b},\, {\bf v})= ({\bf b},\, {\bf u})=0, \qquad
{\bf p}=\xi {\bf b},
$$
where $\xi$ is a scalar, have to be fulfilled.
\vspace*{-12pt}
\paragraph*{Case 2a:}
If the vectors ${\bf v}$ and ${\bf u}$ are not parallel, then 
without loss of generality we may assume that
${\bf b}={\bf u}\times {\bf v}.$  It turns out that in this case $\xi=0$
and ${\bf u}$ and ${\bf v}$
are arbitrary vectors such that $({\bf u},{\bf v})=0.$ The Hamiltonian is
given by
\begin{equation}\label{KOW}
H=\frac{1}{2} \vert {\bf u}\vert ^{2} \vert {\bf M}\vert ^{2}+\frac{1}{2}
\Big({\bf u},\,{\bf M}\Big)^{2}-\frac{\kappa}{2} \Big({\bf v},\,{\bf
M}\Big)^{2}+
\Big({\bf u}\times {\bf v} ,\, {\bf M}\times {\bf \Gamma}\Big).
\end{equation}
The partial integral $P=({\bf u},\,{\bf M})+({\bf v},\,{\bf
\Gamma})$ satisfies the relation
$$
\{H,\, P\}=({\bf u}\times {\bf v},\,{\bf \Gamma})\cdot P.
$$
This Hamiltonian has the following additional integral of fourth degree:
\begin{equation}\label{KOWint}
I=\Big(P \vert {\bf M}\vert ^{2}-2 ({\bf v},\,{\bf M}) ({\bf M},\,{\bf \Gamma})
\Big)\cdot P.
\end{equation}
This integrable case was found in a different non-vector form in \cite{sok,bomasok}. 
A Lax operator is presented in \cite{golsok4}.
\vspace*{-12pt}
\paragraph*{Case 2b:}
The other possibility is that the vectors ${\bf v}$ and ${\bf u}$ are
parallel:
$$
({\bf b},\, {\bf v})=0, \qquad
{\bf p}=\xi {\bf b}, \qquad  {\bf u}=\eta {\bf v}.
$$
It turns out that in this case $\xi=\eta$ and $\kappa=\eta^{2}.$
We see that a real linear integral exists only for the $so(4)$-version
$\kappa > 0$ of bracket (\ref{puas}). The Hamiltonian is given by the formula
\begin{equation}\label{SKL}
H= -2 \eta \Big({\bf v},\,{\bf M}\Big) \Big({\bf z},\,{\bf M}\Big)    +
\Big({\bf b} ,\, {\bf M}\times {\bf \Gamma}\Big), \qquad  \quad
{\bf b}={\bf v}\times {\bf z}
\end{equation}
where ${\bf v}$ and ${\bf z}$ are arbitrary constant vectors.
The partial integral $P=({\bf v},\, \eta {\bf M}+{\bf \Gamma})$ satisfies
$$
\{H,\, P\}=({\bf b},\, \eta {\bf M}+{\bf \Gamma})\cdot P.
$$
This Hamiltonian has the following additional integral of fourth degree (cf. \cite{sklyan,tsig}):
\begin{equation}\label{SKLint}
I=\Big({\bf z},\, (\eta {\bf M}-{\bf \Gamma})  \vert {\bf M}\vert ^{2}+2 {\bf M}\,
({\bf M},\,{\bf \Gamma})
\Big)\cdot P.
\end{equation}
 The eigenvalues $\alpha_i$ of the matrix $A$ from (\ref{HAMHOM})
satisfy the following relations
$$
\alpha_{3}=0, \qquad
\alpha_{1}\alpha_{2}=-\eta^{2}\vert {\bf b}\vert^{2}.
$$
Diagonalizing the matrix $A$, we get the possible canonical form of the Hamiltonian (\ref{SKL}) 
$$
H=\eta (c M_{1}^{2}-\frac{1}{c} M_{2}^{2})+
M_{1} \gamma_{2}-M_{2} \gamma_{1} .
$$
Although it is real only for the $so(4)$-bracket, after the renormalization
$c\eta\rightarrow  c$ of the arbitrary parameter $c$ we get the
Hamiltonian (see \cite{sok2})
$$
H= c  M_{1}^{2}-\frac{\kappa}{c} M_{2}^{2}+
M_{1} \gamma_{2}-M_{2} \gamma_{1},
$$
which is real for any bracket (\ref{puas}). In particular,  if $\kappa=0,$ 
we have a new integrable case on $e(3)$ with a fourth degree integral.

\subsection{Kowalewski-Lyapunov test and a new integrable case.}

In this section we show that the class of Hamiltonians
(\ref{vecGEN}) contains a number of integrable cases. 
The first example of this kind was found in \cite{sok2}:

{\bf Example 1.}  Consider the $so(3,1)$-version
$\kappa < 0$ of bracket (\ref{puas}). 
The Hamiltonian
\begin{equation}\label{case1}
H=\Big({\bf a},\,{\bf b}\Big)\vert {\bf M}\vert ^{2}-
\Big({\bf a},\,{\bf M}\Big)\Big({\bf b},\,{\bf M}\Big)+
\Big({\bf b},\, {\bf M}\times {\bf \Gamma}\Big),
\end{equation}
where the vector ${\bf b}$ is arbitrary and the length of the vector
${\bf a}=(a_1, a_2, a_3)$ is related to the Poisson bracket parameter $\kappa$ by
\begin{equation}\label{norm}
a_{1}^{2}+a_{2}^{2}+a_{3}^{2} =-\kappa,
\end{equation}
possesses the additional quartic integral
$$
I=\Big({\bf b},\,{\bf M}\Big)^{2}
\Big[2 \Big({\bf a},\, {\bf M}\times {\bf \Gamma}\Big)-
\Big({\bf a},\, {\bf M}\Big)^{2} -\kappa
\vert {\bf M}\vert^{2}+\vert {\bf \Gamma}\vert^{2}\Big].
$$

Recently in the paper \cite{tsiggor} the following integrable case has been
found:

{\bf Example 2.} 
The Hamiltonian
\begin{equation}\label{case2}
H=\Big({\bf a},\,{\bf b}\Big)\vert {\bf M}\vert ^{2}-2
\Big({\bf a},\,{\bf M}\Big)\Big({\bf b},\,{\bf M}\Big)+
\Big({\bf b},\, {\bf M}\times {\bf \Gamma}\Big),
\end{equation}
has under condition (\ref{norm}) the additional cubic integral
$$
I=\Big({\bf b},\,{\bf M}\Big)
\Big[2\, \Big({\bf a},\, {\bf M}\times {\bf \Gamma}\Big)-\kappa
\vert {\bf M}\vert^{2}+\vert {\bf \Gamma}\vert^{2}\Big].
$$

In the two examples mentioned above, constraint (\ref{norm}) is necessary for integrability. 
We are going to find all integrable Hamiltonians similar to (\ref{case1})
and (\ref{case2}).
To do that we apply the Kowalewski-Lyapunov test to the class
of Hamiltonians (\ref{vecGEN}) assuming that the additional condition (\ref{norm}) is valid.

Suppose we have a dynamical system 
\begin{equation}\label{sysgen}
\frac{d {\bf X}}{d t}={\bf F}( {\bf X}),  \qquad  {\bf X}=(x_1,\dots , x_N), 
\qquad  {\bf F}=(f_1,\dots f_N)
\end{equation}
where $f_i$ are homogeneous quadratic polynomials of ${\bf X}$. 
Solutions of the form 
\begin{equation}\label{xx}
{\bf X}_0=\frac{\bf K}{t}
\end{equation}
for system (\ref{sysgen}) with ${\bf K}$ being  a constant vector 
are called {\it Kowalewski solutions}.  Substituting (\ref{xx}) into 
(\ref{sysgen}), one obtains a system of algebraic equations for possible 
vectors ${\bf K}.$ 

The linearization ${\bf X}={\bf X_0}+\varepsilon {\bf \Psi}$ of system 
(\ref{sysgen}) on a Kowalewski solution ${\bf X_0}$ obeys
\begin{equation}\label{linsolu}
\frac{d {\bf \Psi}}{d t}=\frac{1}{t} S({\bf \Psi}),
\end{equation}
where $S$ is a constant $N\times N$-matrix depending on the 
Kowalewski solution.

Solutions of (\ref{linsolu}) have the form ${\bf \Psi}={\bf s} \, t^{-k}$,
where $k$ is an eigenvalue and ${\bf s}$ is an eigenvector of the matrix $S.$
The number $1-k$ is called {\it Kowalewski exponent}.

According to the Kowalewski-Lyapunov test, system (\ref{sysgen}) 
is "integrable" if for any Kowa\-lewski solution all corresponding Kowalewski 
exponents belong to an {\it a priori} fixed number set ${\cal A}.$ 
The structure of ${\cal A}$ is closely related to analytic properties
of general solution for (\ref{sysgen}). The usual choice ${\cal A}=\Z$ is
associated with the requirement that the general solution should be 
single-valued. The latter is a standard assumption for the Painleve analysis. 
The most general version ${\cal A}=\Q$ is associated with
general solutions having algebraic branch points. But the main property 
for us is that ${\cal A}$
cannot be too wide. In particular, it cannot contain any open subset of $\C$
or $\R$. Therefore for any one-parameter family of integrable (in 
the Kowalewski-Lyapunov sense) dynamical systems (\ref{sysgen}) 
the Kowalewski exponents must not 
depend on the parameter. This gives us strong necessary integrability 
conditions for one-parametric families of homogeneous quadratic dynamical 
systems.

Given $c_{1}$ and  $c_{2}$, the Hamiltonian  (\ref{vecGEN}) with (\ref{norm})  
depends on one essential continuous parameter. Indeed, the length of ${\bf a}$ is 
fixed by (\ref{norm}), the length of ${\bf b}$ can be normalized by the 
scaling of the Hamiltonian. The transformations
\begin{equation}\label{trans}
{\bf M}\rightarrow T {\bf M}, \qquad {\bf \Gamma}\rightarrow T {\bf \Gamma}
\end{equation}
for any  constant orthogonal matrix $T$ preserve brackets  (\ref{puas}) 
and the form of Hamiltonian (\ref{vecGEN}). Two vectors of a fixed length have 
only one invariant (the angle between them) with respect to 
orthogonal transformation (\ref{trans}).

We want to find all pairs  $c_{1}$ and  $c_{2}$ in (\ref{vecGEN}) such that the Hamiltonian 
is integrable for any value of the angle. 
 
Using transformation (\ref{trans}), we may reduce ${\bf b}$
and ${\bf a}$ to
\begin{equation}\label{normal}
{\bf b}=(0,0,1), \qquad {\bf a}=(a_{1},0, a_{3}).
\end{equation}
Taking into account the constraint (\ref{norm}), we see that now $a_{3}$
remains to be the only free parameter in  (\ref{vecGEN}).
For a generic Hamiltonian of this kind the Kowalewski exponents for the equations of motion depend continuously on $a_{3}$. For an integrable Hamiltonian 
these exponents must not depend on the parameter at all.

{\bf Theorem 1.} Suppose all Kowalewski exponents for Hamiltonian 
(\ref{vecGEN}) with  (\ref{norm}), (\ref{normal}) 
do not depend on $a_{3}$; then the pair of constants $c_{1}, c_{2}$ 
belongs (up to the transformation  $c_{1}\rightarrow -c_{1}, \quad
c_{2}\rightarrow -c_{2}$, which corresponds to 
${\bf a}\rightarrow -{\bf a}$) to the following list \vspace{-6pt}
\begin{tabbing}
\quad \= a) \quad  $c_{1}$ - arbitrary, \ \ \ \ \ \= $c_{2}=\ \ \, 0$;   \\
\> b) \quad  $c_{1}=1,$  \> $c_{2}=-2$;  \\
\> c) \quad  $c_{1}=1,$  \> $c_{2}=-1$;  \\
\> d) \quad  $c_{1}=1,$  \> $c_{2}=-\frac{1}{2}$; \\
\> e) \quad  $c_{1}=1,$  \> $c_{2}=\ \ \, 1$.   \vspace{-12pt}
\end{tabbing}
 \vspace*{-6pt}
To prove this statement, we have to investigate the Kowalewski
exponents on all solutions of the form 
\begin{equation}\label{solu}
 M_{i}=\frac{m_{i}}{t}, \qquad  \gamma_{i}=\frac{g_{i}}{t}
\end{equation}
for the equations of motion defined by Hamiltonian (\ref{vecGEN}), 
 (\ref{norm}).  To get the list of Theorem 1, it turns out to be enough 
to investigate the Kowalewski exponents on special solutions (\ref{solu}), which  
satisfy  $g_{3}=1$. There exist two classes of such solutions. 
For the first class we have $c_{2} (a_{1} m_{2}-a_{2} m_{1})=1$.
The second class is defined by $m_{3}=0.$
For solutions of the first class, besides the cases of Theorem 1, the 
only extra surviving possibility is $c_{1}=1/2, \, c_{2}=-1$. But
this case does not pass the Kowalewski-Lyapunov test on the 
solutions of the second class. For each pair  $(c_1,c_2)$ of the remaining 
list a)-e) of Theorem 1  
we verify that for all Kowalewski solutions the Kowalewski exponents 
do not depend on $a_3$.
All computations are straightforward. A technical problem is that the
explicit form of solutions of the first class involves radicals. To avoid
this difficulty, we used calculations based on the Groebner basis technique.

{\bf Comments.} The Hamiltonian a) belongs to the family
\begin{equation}\label{case0}
H=c_{1} \vert {\bf b}\vert^{2}\,\vert {\bf M}\vert ^{2}+c_{2}
\Big({\bf b},\,{\bf M}\Big)^{2}+
\Big({\bf b},\, {\bf M}\times {\bf \Gamma}\Big),
\end{equation}
which possesses (without restriction (\ref{norm}))
the linear integral of motion $I=({\bf b},\,{\bf M})$.

The Hamiltonians b) and c) are described in Examples 2 and 1,
correspondingly.
The possibility d) leads to the following  new integrable case with an
additional integral of sixth degree:

{\bf Example 3.} The Hamiltonian
\begin{equation}\label{case3}
H=\Big({\bf a},\,{\bf b}\Big)\vert {\bf M}\vert ^{2}-\frac{1}{2}
\Big({\bf a},\,{\bf M}\Big)\Big({\bf b},\,{\bf M}\Big)+
\Big({\bf b},\, {\bf M}\times {\bf \Gamma}\Big),
\end{equation}
under condition (\ref{norm}) has the additional sixth degree integral
$$
I=\Big({\bf b},{\bf M}\Big)^{2}\Big[\Big({\bf b}\times {\bf a},
{\bf M}\times {\bf a} \Big)^{2}\,{\bf M}^{2}+
2 \Big({\bf b}\times {\bf a},
{\bf M}\times {\bf a} \Big)\Big({\bf b}\times{\bf a},
{\bf M}\times({\bf M}\times {\bf \Gamma})\Big)
$$
$$
-{\bf \Gamma}^{2}\Big({\bf M},{\bf b}\times{\bf a}\Big)^{2}-
\Big({\bf b}\times{\bf a},
{\bf M}\times {\bf \Gamma}\Big)^{2}-{\bf M}^{2}{\bf \Gamma}^{2}\Big({\bf b},{\bf
a}\Big)^{2}-\kappa {\bf M}^{2}{\bf \Gamma}^{2} {\bf b}^{2}
 \Big].
$$
Possibly this integrable case is related to the simple Lie algebra $g(2)$.

In \cite{soktsig1} a generalization of the general scheme by Sklyanin
\cite{sklyan} has been proposed.
In  \cite{tsiggor}  the authors found a separation of variables for
Hamiltonian (\ref{case2}) from Example 2 in the framework of this approach.
Probably a separation of variables for the Hamiltonians from 
Examples 1 and 3 could be found after some development of these ideas.  

All Kowalewski exponents for cases a)-c) are integers. In case d) there are
two solutions of the class 2. For these solutions we have
$$
\hbox{det}(S-k \,\mbox{Id})=(k-1)^{3}(k+2)(k-\frac{1}{2})(k+\frac{3}{2})
$$
and
$$
\hbox{det}(S-k \,\mbox{Id})=(k-1)^{3}(k+2)(k+\frac{1}{2})^{2}.
$$
In other words, some of the Kowalewski exponents are half-integers.

Case e) is a mysterious one. We have verified that the Hamiltonian has no
polynomial additional integrals of degrees not greater than 8. On the
other hand, on all Kowalewski solutions all Kowalewski exponents are integers.
It would be interesting to verify whether the equations of motion in the 
case e) satisfy the standard Painleve test.

\section{Inhomogeneous integrable Hamiltonians.}

\subsection{Admissible linear terms for integrable homogeneous 
Hamiltonians}

In this subsection we find for integrable  homogeneous Hamiltonians $H$ of 
the form (\ref{case0}), (\ref{case2}), (\ref{case1}), (\ref{SKL}), 
(\ref{KOW}) or (\ref{case3}) possible linear terms 
\begin{equation}\label{tail}
T=({\bf k},\, {\bf M})+({\bf n},\, {\bf \Gamma}),
\end{equation}
where ${\bf k}$ and ${\bf n}$ are constant vectors, such that the Hamiltonian 
$$
\tilde H=H+T
$$ 
has an additional integral of the same degree as $H$.

{\bf Proposition 1.} The following linear terms are admissible 
(in the above sense):
\begin{tabbing}
5) \quad \= for Hamiltonian (\ref{KOW}) with $({\bf u},{\bf v})=0$ \ \ \ \ \=
$T=p_1 ({\bf u},\, {\bf M})+p_2 ({\bf u}\times {\bf v},\, {\bf \Gamma})$;
\kill 
1) \> for Hamiltonian (\ref{case0}) 
   \> $T=p_1 ({\bf b},\, {\bf M})+p_2 ({\bf b},\, {\bf \Gamma})$; \\ \> \> \\
2) \> for Hamiltonian (\ref{case2}) with  (\ref{norm}) 
   \> $T=({\bf k},\, {\bf M})+ p_1 ({\bf b},\, {\bf \Gamma})$; \\ \> \> \\
3) \> for Hamiltonian (\ref{case1}) with  (\ref{norm})
   \> $T=(p_1 {\bf a}+p_2 {\bf a}\times {\bf b},\, {\bf M})+ 
      p_3 ({\bf b},\, {\bf \Gamma})$;  \\ \> \> \\
4) \> for Hamiltonian  (\ref{SKL})
   \> $T=({\bf k},\, {\bf M})+ p_1 ({\bf v}\times {\bf z},\, {\bf \Gamma})$;
      \\ \> \> \\
5) \> for Hamiltonian (\ref{KOW}) with $({\bf u},{\bf v})=0$ 
   \> $T=p_1 ({\bf u},\, {\bf M})+p_2 ({\bf u}\times {\bf v},\, {\bf \Gamma})$;
      \\ \> \> \\
6) \> for Hamiltonian (\ref{case3}) with  (\ref{norm})
   \> $T=p_1 ({\bf a}\times {\bf b},\, {\bf M})$,
\end{tabbing}
where ${\bf k}$ is an arbitrary vector and $p_1, p_2, p_3$ 
are arbitrary constants.

We 
present the explicit form of the additional integrals for the  
non-homogeneous Hamiltonians $\tilde H$ of Proposition 1 
in appendix A. 

\subsection{A deformation of the Poincare model}
The formula (\ref{case0}) describes Hamiltonians (\ref{genham}) that have
linear additional integrals. We shall call (\ref{case0}) the Poincare model.

For the 
special pair $c_{1}=1, c_{2}=-1/2$ the Poincare model 
is superintegrable. It means that besides 
the linear integral the Hamiltonian has a polynomial
integral of degree higher than 1.

{\bf Proposition 2.} The Hamiltonian 
\begin{equation}\label{superham}
H_{\mbox{\scriptsize hom}}=\vert {\bf b}\vert^2\,\vert {\bf M}\vert ^{2}-\frac{1}{2}
\Big({\bf b},\,{\bf M}\Big)^2+
\Big({\bf b},\, {\bf M}\times {\bf \Gamma}\Big)
\end{equation}
under condition $\vert {\bf b}\vert ^{2}=-\kappa$ 
has the following additional integral of degree 4, functionally 
independent of $H$, the Casimirs and the linear integral 
$({\bf b},\, {\bf M})$: 
\begin{eqnarray*}
I&=&
({\bf k},\, {\bf M}) \left[({\bf k},\, {\bf M})\vert {\bf b}\vert^2 -2 
({\bf k},\, {\bf b}) ({\bf b},\, {\bf M}) \right]\cdot 
\left[\vert {\bf \Gamma}\vert^2+\vert {\bf b}\vert^2
\vert {\bf M}\vert^2 \right] \\
&&
+\vert {\bf M}\vert^2 ({\bf b},\, {\bf M})^2 \left[ ({\bf k},\, {\bf b})^2+
\vert {\bf k}\vert^2 \vert {\bf b}\vert^2 \right]-\Big(
{\bf k}\times {\bf b},\,{\bf M}\times {\bf \Gamma} \Big)^2 \\
&&
+2 \Big(
{\bf k}\times {\bf b},\,{\bf b}\times {\bf M}  \Big)\cdot 
\left[\vert {\bf M}\vert^2 ({\bf k},{\bf b}\times{\bf \Gamma})
-({\bf M},\,{\bf \Gamma})({\bf k},{\bf b}\times{\bf M})
\right], 
\end{eqnarray*}
where ${\bf k}$ is an arbitrary constant vector.

It follows from the Jacobi identity that $I_1=
\{({\bf b}, {\bf M}),\, I \}$ is a first integral as well. 
It turns out that $I_1\ne 0 ,$ which means that the Poisson subalgebra of 
polynomial integrals for $H_{\mbox{\scriptsize hom}}$ is non-commutative. For any $X$ 
let us denote the Poisson bracket $\{({\bf b},{\bf M}),\, X\}$ by $X'$. 
One can check that the integral $I$ satisfies the relation 
$I'''+4 \vert {\bf b}\vert^2 I'=0.$ 

The derivation $X \rightarrow X'$ can be 
regarded as a linear operator on the finite dimensional 
vector space $S_n$ of all homogeneous polynomials of degree 
$n$ depending on components of ${\bf M}$ and ${\bf \Gamma}.$ The operator 
spectrum is $\mu_k=i k \vert {\bf b}\vert, \, 1\le k \le n. $
Probably this operator plays an important role in the theory of the vector 
Hamiltonians (\ref{vecGEN}). Many terms in the integrable Hamiltonians 
from this class admit simple descriptions in terms of this operator. 
For example, the linear polynomial 
$\spa{k}{b}{M}$, where ${\bf k}$ is an arbitrary 
constant vector, is the general linear solution of the
equation $X''+ \vert {\bf b}\vert^2 X=0.$ This fact and formula (\ref{pin}) 
imply that the arbitrary Hamiltonian (\ref{vecGEN}) satisfies the equation
$X'''+\vert {\bf b}\vert^2 X'=0.$ The additional integral from Example 1 
satisfies the same equation and so on. 

The Hamiltonian (\ref{superham}) admits the following 
inhomogeneous integrable extension:
\begin{equation}\label{lintail}
\tilde H=H_{\mbox{\scriptsize hom}}+
({\bf k}\times {\bf b},\, {\bf M})+p_1 ({\bf b},\, {\bf \Gamma}),
\end{equation}
where ${\bf k}$ is an arbitrary constant vector, $p_1$ is an arbitrary constant. 
Hamiltonian (\ref{lintail}) under condition (\ref{norm}) has an additional 
integral of degree 4, given in the appendix A.  

\section{Classification results}

It is very likely that all integrable Hamiltonians of the form 
(\ref{genham}) and (\ref{genhamnon}) are
exhausted by the examples presented in sections 2 and 3.

{\bf Theorem 2.} Suppose a Hamiltonian of the form (\ref{genham}) with real 
coefficients has an
additional polynomial integral of degree from 1 to 8; then the Hamiltonian
belongs to the six families 
(\ref{case0}), (\ref{case2}), (\ref{case1}), (\ref{SKL}), (\ref{KOW})  
or (\ref{case3}). 

{\bf Scheme of the proof.} Using transformations (\ref{trans}), one can
reduce any (real) Hamiltonian (\ref{genham}) to
\begin{equation}\label{redham}
H=a_{1}M_{1}^{2}+a_{2} M_{2}^{2}+a_{3}M_{3}^{2}+ a_{4} M_{1} M_{3}+
 a_{5} M_{2} M_{3}+M_{1}\gamma_{2}-M_{2}\gamma_{1}.
\end{equation}
In this canonical form the vector ${\bf b}$ is normalized to $(0,0,1)$. 
The alternative idea of bringing the matrix $A$ to the diagonal form
lead to overwhelming computational complexity.

Without loss of generality we may assume that the additional
polynomial integral is homogeneous. Given the degree $m$ of the
additional integral $I$, we form the general homogeneous $m$-th degree
polynomial of six variables $M_i, \gamma_i$ with undetermined
coefficients. The condition $\{I,\, H \}=0$ gives rise to a bi-linear
system of algebraic equations for both coefficients of $H$ and $I$. Of
course, this system can be solved ``by hand'' only for small $m$.  If
$m=6$ the algebraic system contains 791 bi-linear equations for 458 unknown
coefficients. This system can  not be currently solved by
standard computer algebra systems. 
Also all attempts by the authors to use the two best known 
packages specialized in the
solution of polynomial algebraic systems failed. 

The computation was performed using the computer algebra package {\sc
Crack}.  For $m<6$ the calculations were performed automatically and for
$m\geq 6$ with manual interaction.

Following the same line we have obtained  

{\bf Theorem 3.} Suppose a Hamiltonian of the form (\ref{genhamnon}) 
with real coefficients has an
additional polynomial integral of degree from 1 to 6; then the Hamiltonian
belongs to the seven families described in Propositions 1 and 2. 

\section{Conclusion: unsolved problems}
This paper as well as \cite{sok2} belongs mostly to so called ``experimental'' 
mathematical physics. The result of the experiment is a new interesting class 
(\ref{genhamnon}) of quadratic Hamiltonians. This class contains several new 
integrable cases. Theorems 1-3 give reasons to believe that we found all 
real integrable Hamiltonians of the form (\ref{genhamnon}). However, this  
should be proved. To do that, one can apply the Painleve approach or methods 
developed in \cite{ziglin}. 

The separation of variables for several models from our paper is also 
an open problem. 

Besides Hamiltonians listed in Theorem 3, there exist integrable 
Hamiltonians of the form (\ref{genhamnon}) with complex coefficients. 
For example, the Hamiltonian (\ref{vecGEN}), (\ref{norm}) with $c_1=1, 
c_2=-\frac{2}{3}$ has an additional integral of sixth degree under condition 
$\vert {\bf b}\times {\bf a}\vert=0.$ To find the complex Hamiltonians, 
one should consider two different normalizations of the vector ${\bf b}$. 
The first is  ${\bf b}=(0,0,b_3)$, which corresponds to the possibility 
$\vert {\bf b}\vert \ne 0$ and gives rise to the normal form 
(\ref{redham}). The second normalization ${\bf b}=(0,b_2,i\, b_2)$ 
corresponds to $\vert {\bf b}\vert=0.$ In the case of fourth degree 
additional integrals all complex integrable Hamiltonians have been found in 
\cite{sok2}.

Probably all Hamiltonians from our paper have their quantum counterparts 
(for the case of fourth degree integral see \cite{sok2}). It would be 
interesting to find the corresponding quantum Hamiltonians and integrals 
as well as the quantum separation of variables.

\section*{Acknowledgments} 
The authors are grateful to A.V.\ Tsiganov for 
useful discussions and Winfried Neun for computer algebra systems support.
The first author (V.S.) is grateful to Brock University and the 
Fields Institute for hospitality. The research was partially supported 
by RFBR grants 02-01-00431 and NSh 1716.2003.1 (for V.S.) and NSERC 
research grant 253296 (for T.W.).

\section*{Appendix A} 
Here we present an explicit form of additional integrals for inhomogeneous 
Hamiltonians from Propositions 1 and 2. Each integral I of degree $m$ 
is a sum 
$
I=\sum_{i=1}^m I_i,
$ 
where $I_i$ is a polynomial of degree $i$ homogeneous in 
${\bf M}, {\bf \Gamma}$. 

\noindent{\bf Case 1 of Proposition 1.}  $I=I_1=({\bf b}, {\bf M})$.       

\noindent{\bf Case 2 of Proposition 1.}  $I_3$ is presented in Example 2, 
\begin{eqnarray*}
I_2&=&2 p_1 ({\bf b},{\bf M}) ({\bf a},{\bf \Gamma})-
({\bf k},{\bf a}) \vert {\bf M}\vert^2-\spa{k}{M}{\Gamma}, \\
I_1&=&-\kappa p_1^2 ({\bf b},{\bf M})-p_1 ({\bf k},{\bf \Gamma}). 
\end{eqnarray*}

\noindent{\bf Case 3 of Proposition 1.}  $I_4$ is presented in Example 1,
\begin{eqnarray*}
I_3&=&2 ({\bf b},{\bf M}) \,\left\{p_1\Big[({\bf a},{\bf M})^2 +
\kappa \vert {\bf M}\vert^2-
 \vert {\bf \Gamma}\vert^2-2 \spa{a}{M}{\Gamma}
 \Big]\right.  \\
&&
+ \left.  p_2 \Big[
({\bf b}\times {\bf a},\, {\bf M}\times {\bf \Gamma})+ ({\bf a},{\bf M})
 \spa{a}{b}{M}
 \Big]+  p_3  ({\bf b},{\bf M})({\bf a},{\bf \Gamma})\right\}, 
\\ &&\\
I_2&=& p_1^2\,\Big[2 \vert {\bf \Gamma}\vert^2-({\bf a},{\bf M})^2 + 
2 \spa{a}{M}{\Gamma} \Big]
- 2 p_2 p_3 ({\bf b},{\bf M}) \spa{a}{b}{\Gamma} \\
&& 
+ p_2^2\,\Big[ \vert {\bf b}\vert^2 ({\bf a},{\bf M})^2-
\kappa  ({\bf b},{\bf M})^2 -2 ({\bf a},{\bf b})
({\bf a},{\bf M})({\bf b},{\bf M}) \Big] 
- 4 p_1 p_3 ({\bf b},{\bf M}) ({\bf a},{\bf \Gamma}) \\ 
&&
- p_3^2 \kappa ({\bf b},{\bf M})^2+ 2 p_1 p_2 
\Big[({\bf a}\times {\bf b},\, {\bf M}\times {\bf \Gamma})
- ({\bf a},{\bf M}) 
 \spa{a}{b}{M} \Big]\\ &&\\
I_1&=& 2 p_1 p_3 \Big[p_1 ({\bf a},{\bf \Gamma})+
p_2 \spa{a}{b}{\Gamma}+p_3 \kappa ({\bf b},{\bf M})\Big].
\end{eqnarray*}

\noindent{\bf Case 4 of Proposition 1.}  
$I_4=\eta\vert{\bf a}\times{\bf b}\vert^ 2I$ 
where $I$ is given by (\ref{SKLint}),
\begin{eqnarray*}
I_3&=&p_1 \eta \Big( ({\bf v},\, {\bf z})^2 -
 \vert {\bf v} \vert^2 \vert {\bf z} \vert^2 \Big)\cdot
\Big[\eta \vert {\bf M} \vert^2  ({\bf v},{\bf z}\times {\bf \Gamma} )-
2  ({\bf v},\, {\bf \Gamma})  ({\bf z},{\bf M}\times {\bf \Gamma} )+
\vert {\bf \Gamma} \vert^2  ({\bf v},{\bf z}\times {\bf M} ) \Big] \\
&&
+\eta^2  \vert {\bf M} \vert^2 \Big[ ({\bf z},\, {\bf M})
 ({\bf v}\times {\bf z},\,{\bf k}\times {\bf v} ) -  ({\bf v},\, {\bf M})
 ({\bf v}\times {\bf z},\,{\bf k}\times {\bf z} )  \Big] \\
&&
+2 \eta \Big[  ({\bf v},\, {\bf M}) ({\bf z},\, {\bf k})
 ({\bf v}\times {\bf z},\,{\bf \Gamma}\times {\bf M} )+ 
({\bf v},\, {\bf M}) ({\bf k},\, {\bf M})
 ({\bf v}\times {\bf z},\,{\bf z}\times {\bf \Gamma} ) \\
&& - ({\bf v},\, {\bf M}) ({\bf k},\, {\bf \Gamma})
 ({\bf v}\times {\bf z},\,{\bf z}\times {\bf M} ) \\
&&
-\frac{1}{2}  ({\bf k},\, {\bf \Gamma})  \vert {\bf M} \vert^2 
 \vert{\bf v}\times {\bf z}\vert^2- 
({\bf z}\times {\bf M},\,{\bf M}\times {\bf \Gamma} ) 
 ({\bf v}\times {\bf z},\,{\bf v}\times {\bf k} ) \Big] \\
&&
- ({\bf v},{\bf z}\times {\bf k} )  ({\bf v},{\bf z}\times {\bf M} ) 
 \vert {\bf \Gamma} \vert^2,\\&&\\
I_2&=&p_1^2 \eta  \vert {\bf v}\times {\bf z} \vert^2 \Big[\eta^2 
 ({\bf v}\times {\bf M},\,{\bf z}\times {\bf M} )+
 ({\bf v},\, {\bf \Gamma})  ({\bf z},\, {\bf \Gamma}) \Big]
-p_1 \eta \Big[ \Big(({\bf v},{\bf M}) ({\bf z},{\bf \Gamma})\\
&&
 + ({\bf v},{\bf \Gamma}) ({\bf z},{\bf M})
\Big) ({\bf v},{\bf z}\times {\bf k} ) 
+ ({\bf v}\times {\bf z},\,{\bf z}\times {\bf k} ) 
 ({\bf v},{\bf M}\times {\bf \Gamma} ) \\
&&
+ ({\bf v}\times {\bf z},\,{\bf v}\times {\bf k} ) 
 ({\bf z},{\bf M}\times {\bf \Gamma} )\Big] 
-\eta  \vert {\bf M} \vert^2 
 ({\bf v}\times {\bf k},\,{\bf z}\times {\bf k} )+ 
 ({\bf v},{\bf z}\times {\bf k} ) ({\bf k},{\bf M}\times {\bf \Gamma} ),\\&&\\
I_1&=&p_1 ({\bf v},{\bf z}\times {\bf k} ) \cdot \Big(p_1 \eta^2 
 ({\bf v},{\bf z}\times {\bf M} ) +
({\bf k},\,{\bf \Gamma}) \Big).
\end{eqnarray*}

\noindent{\bf Case 5 of Proposition 1.}  $I_4$ is is given by (\ref{KOWint}),  $I_1=0$,  
\begin{eqnarray*}
I_3&=&2 p_1 \Big( P \vert{\bf M}\vert^2-
({\bf v},{\bf M}) ({\bf M},{\bf \Gamma}) \Big)+
2 p_2 \Big( ({\bf M},{\bf \Gamma}) \spa{u}{v}{M}-P \spa{M}{\Gamma}{v} \Big),\\
I_2&=&p_1^2 \vert{\bf M}\vert^2+p_2^2 \Big( \vert{\bf v}\vert^2 
\vert{\bf \Gamma}\vert^2+\kappa ({\bf v},{\bf M})^2
- ({\bf v},{\bf \Gamma})^2 \Big) - 2 p_1 p_2 \spa{M}{\Gamma}{v}, 
\end{eqnarray*}
where $P=({\bf u},{\bf M})+({\bf v},{\bf \Gamma})$.

\noindent{\bf Case 6 of Proposition 1.}  $I_6$ is presented in Example 3, \, 
$I_1=0$,  
\begin{eqnarray*}
I_5&=&4 p_1({\bf b},{\bf M})\left\{\Big[ ({\bf a},{\bf b})  \vert{\bf M}\vert^2 
 ({\bf a}\times {\bf b},\,{\bf a}\times {\bf M})\, \spa{a}{b}{M}+
 \vert{\bf \Gamma}\vert^2 
 ({\bf a}\times {\bf b},\,{\bf b}\times {\bf M})\, \spa{a}{b}{M}
  \Big] \right. \\
&&
+ ({\bf a}\times {\bf b},\,{\bf M}\times {\bf \Gamma})\, \Big[ ({\bf b},{\bf \Gamma})  \spa{a}{b}{M} - 
({\bf b},{\bf M})  \spa{a}{b}{\Gamma} 
 \Big] \\
&&
+ ({\bf b},{\bf M}) ({\bf M},{\bf \Gamma}) \Big[ 
 ({\bf a},\,{\bf b})    ({\bf a}\times {\bf b},\,{\bf a}\times {\bf M })
 -\kappa   ({\bf a}\times {\bf b},\,{\bf b}\times {\bf M })
  \Big] \\
&&
- ({\bf a},\,{\bf b}) ({\bf a}\times {\bf b},\,{\bf b}\times {\bf M }) \Big[ 
 2  ({\bf a},\,{\bf M}) ({\bf M},\,{\bf \Gamma}) 
- ({\bf a},\,{\bf \Gamma}) \vert{\bf M}\vert^2 
  \Big] \\
&&
+ \left. \vert{\bf M}\vert^2    ({\bf a}\times {\bf b},\,{\bf a}\times {\bf M }) \Big[ 
 \vert{\bf b}\vert^2 ({\bf a},\,{\bf \Gamma}) - 2  ({\bf a},\,{\bf b}) ({\bf b},\,{\bf \Gamma}) 
  \Big]\right\},\\ &&\\
I_4&=&p_1^2\left\{ 4 \Big[({\bf a}\times {\bf b},\, {\bf b}\times {\bf M})\,
\Big(({\bf a},\,{\bf \Gamma}) ({\bf b},\,{\bf M})+2 
({\bf a},\,{\bf b}) ({\bf M},\,{\bf \Gamma})
 \Big)-  ({\bf b},\,{\bf \Gamma}) ({\bf b},\,{\bf M})\,
({\bf a}\times {\bf b},\, {\bf a}\times {\bf M}) \right. \\
&&
- 2  ({\bf a},\,{\bf b}) \vert {\bf M}\vert^2 
 ({\bf a}\times {\bf b},\, {\bf b}\times {\bf \Gamma}) \Big]
\cdot \spa{a}{b}{M} +4 ({\bf b},\,{\bf M}) 
\Big[ ({\bf b},\,{\bf M}) 
 ({\bf a}\times {\bf b},\,{\bf a}\times {\bf M}) \\
&&
- ({\bf a},\,{\bf M}) 
  ({\bf a}\times {\bf b},\,{\bf b}\times {\bf M})   \Big]
\cdot \spa{a}{b}{\Gamma}-8 
 ({\bf a}\times {\bf b},\,{\bf b}\times {\bf M})
 ({\bf a}\times {\bf b},\,{\bf b}\times {\bf \Gamma})
 ({\bf M},\,{\bf \Gamma})  \\
&&
+ \Big[4 ({\bf a},\,{\bf b})^2 ({\bf a},\,{\bf M}) 
 ({\bf a}\times {\bf b},\,{\bf b}\times {\bf M})-4 
({\bf b},\,{\bf M}) \Big(2  ({\bf a},\,{\bf b})^2- 
 \vert {\bf a}\vert^2  \vert {\bf b}\vert^2    \Big)
({\bf a}\times {\bf b},\,{\bf a}\times {\bf M}) \\
&&
+ \left. 4 ({\bf a}\times {\bf b},\,{\bf b}\times {\bf \Gamma})^2\Big]
  \vert{\bf M}\vert^2
-  4 ({\bf a},\,{\bf b})^2 \Big( ({\bf a},\,{\bf b})^2- 
 \vert {\bf a}\vert^2  \vert {\bf b}\vert^2  \Big) \vert {\bf M}\vert^4\right\},\\&&\\
I_3&=&8 p_1^3\vert {\bf a}\times {\bf b}\vert^2 
\Big(({\bf a},{\bf b}) \vert{\bf M}\vert^2 \spa{a}{b}{M} +  
 ({\bf b}\times {\bf a},\,{\bf b}\times {\bf \Gamma}) \vert{\bf M}\vert^2+
 ({\bf a}\times {\bf b},\,{\bf b}\times {\bf M}) ({\bf M},{\bf \Gamma})
  \Big),\\&&\\
I_2&=&-4 p_1^4\vert {\bf a}\times {\bf b}\vert^2 
\Big(({\bf a},{\bf b})^2  \vert{\bf M}\vert^2 -  
 \vert{\bf b}\vert^2 \vert{\bf \Gamma}\vert^2 \Big).
\end{eqnarray*}

\noindent{\bf The Case of Proposition 2.} $I_1=0$; $I_4$ is related to integral 
$I$ from Proposition 2 by
\begin{eqnarray*}
I_4&=&\vert {\bf b}\vert^2 I+({\bf b}, {\bf M})^2 
\Big[({\bf k}\times {\bf b},\,{\bf b}\times {\bf \Gamma})\, \spa{k}{b}{M} - 
({\bf k}\times {\bf b},\,{\bf b}\times {\bf M})\, \spa{k}{b}{\Gamma} \\
&&
+\Big(({\bf k},\,{\bf b})^2-\vert {\bf k}\vert^2 \vert {\bf b}\vert^2\Big)
 \cdot \Big(\vert {\bf \Gamma}\vert^2 - \frac{1}{4}({\bf b}, {\bf M})^2
\Big)+\vert {\bf k}\vert^2 \vert {\bf b}\vert^2 
 \Big( \vert {\bf \Gamma}\vert^2+\kappa  \vert {\bf M}\vert^2 \Big)
 \Big],\\&&\\
I_3&=&p_1 \Big[\Big(({\bf k},\,{\bf b})^2
-\vert {\bf k}\vert^2 \vert {\bf b}\vert^2\Big)
({\bf b}, {\bf M})^2 ({\bf b}, {\bf \Gamma})
-2  \vert {\bf b}\vert^2\, 
({\bf k}\times {\bf b},\,{\bf M}\times {\bf \Gamma})\,
\Big(\spa{k}{b}{\Gamma}+({\bf k}\times {\bf b},\,{\bf b}\times {\bf M})  \Big)\Big]\\
&&
+\Big(({\bf k},\,{\bf b})^2-\vert {\bf k}\vert^2 \vert {\bf b}\vert^2\Big)
\,({\bf b}, {\bf M})\,\Big[({\bf b}, {\bf M})\,\spa{k}{b}{M}+
2  ({\bf k}\times {\bf b},\,{\bf \Gamma}\times {\bf M})   \Big],\\&&\\
I_2&=&p_1^2 \Big[\vert {\bf b}\vert^2
 ({\bf k}\times {\bf b},\,{\bf b}\times {\bf M})^2+
({\bf k}\times {\bf b},\,{\bf b}\times {\bf \Gamma})^2 +
(({\bf k},\,{\bf b})^2-\vert {\bf k}\vert^2 \vert {\bf b}\vert^2)
\vert {\bf b}\times {\bf \Gamma}\vert^2 \\
&&
+ \vert {\bf k}\vert^2 \vert {\bf b}\vert^4 ( \vert {\bf \Gamma}\vert^2+\kappa 
 \vert {\bf M}\vert^2) \Big] 
-2 p_1 \Big(({\bf k},\,{\bf b})^2-\vert {\bf k}\vert^2 \vert {\bf b}\vert^2\Big) \, 
({\bf b}, {\bf M})\,\spa{k}{b}{\Gamma}+\\
&&
\Big(({\bf k},\,{\bf b})^2-\vert {\bf k}\vert^2 \vert {\bf b}\vert^2\Big) 
\Big[({\bf b}, {\bf M})\, 
({\bf k}\times {\bf b},\,{\bf k}\times {\bf M})- 
({\bf k}, {\bf M})\, 
({\bf k}\times {\bf b},\,{\bf b}\times {\bf M})
  \Big].
\end{eqnarray*}

\section*{Appendix B} 
With the comments in this appendix we only want to give some
impression on features of the package {\sc Crack} that we used for the
solution of the overdetermined algebraic systems $\{I,\, H \}=0$. 
The description is neither complete not detailed.

\begin{description} 
  \item[1.] Flexibility (different possibilities to change or 
  adapt the strategy):

  \begin{description}

    \item[a)] {\tt CRACK} has a priority list, which contains among
    other things, 10 different types of substitutions, two types of
    factorizations, three types of shortening of equations, various
    types of Groebner basis calculation steps and the possibility to
    call external specialized packages.

    One can set the different priorities for these internal procedures
    before the session but also change them interactively during the
    computation. In addition, the program is able to make minor changes in
    the priority list in accordance to its analysis of the current
    situation and history of the session. An important point for the
    analysis is that each equation has a property list describing the
    previous use of the equation as well as its potential for further
    computation.

    The program has automatic and interactive modes and the possibility to
    switch between both during the computation. In the interactive mode it
    provides extended support for inspecting the system of equations, to
    specify the very next steps in detail or modify the solution strategy
    in general.

    \item[b)] After a decision is 
    made what kind of step should come next (based
    on the priority list) more heuristic knowledge decides how to do it.
    For example, if a factorization is to come next then a
    refined weighting scheme determines which equation to factorize 
    and in which order to set its factors to zero. 

    \item[c)] During computation inequalities are actively gathered
    and simplified in order to reduce the number of branches.

  \end{description}

  \item[2.] The program is paranoid about expression swell and for its 
  prevention it is able to 
  \begin{description}

    \item[a)] compute a tight upper bound for the system's length
    after an intended substitution and hence choose substitutions
    giving minimal growth,

    \item[b)] find linear combinations of pairs of equations in an
    attempt to shorten and simplify the system recursively during
    computation, and 

    \item[c)] it takes advantage of the initial bi-linear form of the
    system and keeps it linear in the unknown coefficients of the
    first integral at all times.

  \end{description}

  \item[3.] Several facilities are provided, e.g.\ 
  \begin{description}

    \item[a)] to find re-parametrizations of solutions in order to
    merge them,

    \item[b)] safety precautions: to interrupt intermediate steps
    automatically if they (unexpectedly) take too much time, to catch
    all interactive input, to conveniently store and load backups,

    \item[c)] an automatic preparation of web pages with results of the
    computation.

  \end{description}

  \item[4.] Different branches of the general solution can be
  investigated at the same time in parallel on a cluster of computers.

\end{description}

\noindent  To obtain a free copy of this REDUCE package contact T.\ Wolf
under {\tt twolf@brocku.ca} .

\end{document}